\documentclass[twocolumn,showpacs,preprintnumbers,amsmath,amssymb]{revtex4}
\usepackage{amssymb}
\usepackage{graphicx,color}
\usepackage{bm}

\begin{document}

\title{Refraction of Gluon jet in QGP}

\author{Hanzhong Zhang  and  Enke Wang}

\affiliation{Institute of Particle Physics, Huazhong Normal
University, Wuhan 430079, China\\
Key Laboratory of Quark $\&$ Lepton Physics (Huzhong Normal
University), Ministry of Education, China}

\date{\today}

\begin{abstract}
We introduce photon-like refraction for gluon jet in quark gluon
plasma created in high energy nucleus-nucleus collisions. A quark
jet goes straightly while a gluon jet is assumed to bend in the
non-homogeneous medium. Combining refraction and energy loss, we
calculate the away-side dihadron $\Delta\phi$ distribution in
central $Au+Au$ collisions at RHIC energy. The dihadrons from
quark-quark pair center around $\Delta\phi=\pi$, while the
dihadrons from quark-gluon and gluon-gluon pair are deflected from
$\Delta\phi=\pi$. The resulting distribution shows three peaks at
intermediate $p_T$ since both quark-quark and quark-gluon are
important. However we only observe one center peak at high $p_T$,
where the quark-quark contribution dominates. These observation
qualitatively agrees with the experiment data.

\end{abstract}

\pacs{12.38.Mh, 24.85.+p; 25.75.-q}

\maketitle

A strongly couple partonic matter (sQGP) is now believed to exist
in high energy nucleus-nucleus collisions with an extended volume
and time. The high transverse momentum $p_T$ partonic jets
traversing the hot dense matter lose a significant amount of their
energy via induced gluon radiations, a phenomena known as
jet-quenching~\cite{Gyulassy:2003mc}. Such energy loss is predicted
to lead to strong suppression of both single- and correlated
away-side dihadron spectra at high $p_T$
~\cite{xnw04,jiajy05,zoww07}, consistent with experimental
findings~\cite{phenix-star-single,star03-06,phenix-jjia07}.

The suppression factor of the leading hadrons from jet fragmentation
depends on the total parton energy loss which in turn is related to
the jet propagation path weighted with the gluon density
$\rho_g$~\cite{xnw04,zoww07},

\begin{equation}
\Delta E\approx \langle \frac{dE}{dL}\rangle_{1d}
\int_{\tau_0}^{\infty} d\tau \frac{\tau-\tau_0}{\tau_0\rho_0}
\rho_g(\tau,{\bf b},{\bf r}+{\bf n}\tau), \label{eqn:enloss}
\end{equation}
for a parton produced at a transverse position ${\bf r}$ and
traveling along the direction ${\bf n}$. The energy dependence of
the energy loss is parameterized as

\begin{equation}
 \langle\frac{dE}{dL}\rangle_{1d}=\epsilon_0 (E/\mu_0-1.6)^{1.2}
 /(7.5+E/\mu_0),
\label{eqn:loss}
\end{equation}
from the numerical results in Ref.~\cite{ww} in which thermal gluon
absorption is also taken into account. The parameter $\epsilon_0$
should be proportional to the initial gluon density. It was shown
that the single hadrons at high $p_T$ are dominated by the jets
emitted from surface layer of the overlap. However, the away-side
dihadrons contains contribution from both tangentially emitted or
punch-through dijets.
A simultaneous fit to the single and dihadron data constrains the
energy loss parameter within a narrow range: $\epsilon_0=1.6-2.1$
GeV/fm~\cite{zoww07}.

Recently the azimuthal angle $\Delta\phi$ correlations are presented
\cite{phenix-jjia07} for charged hadrons from dijets for
$0.4<p_T<15$~GeV/$c$ in Au+Au collisions at $\sqrt{s_{\rm{NN}}}$ =
200 GeV by PHENIX group. Especially, with increasing $p_T$, the
away-side distribution evolves from a broad to a concave shape, then
to a convex shape \cite{phenix-jjia07}. Comparisons to $p+p$ data
suggest that the away-side can be divided into a partially
suppressed ``head'' region centered at $\Delta\phi\sim\pi$, and an
enhanced ``shoulder'' region centered at $\Delta\phi\sim \pi\pm1.1$.
These away-side results are then used to constrain various proposed
mechanisms, such as mach-cone~\cite{Casalderrey-Solana:2004qm},
large angle gluon radiation~\cite{Vitev:2005yg,Polosa:2006hb},
Cherenkov gluon radiation~\cite{Koch:2005sx} or deflected
jets~\cite{Armesto:2004pt,Chiu:2006pu}. The goal of current letter
is to propose gluon refraction as yet another mechanism for the
away-side behavior, based on a NLO pQCD parton model with jet
quenching in $A+A$ collisions.

For an optical ray propagating in non-homogeneous medium with a
refractive index distribution $n(\bf r)$. Its pathway in the medium
should satisfies the relation $\delta \int n({\bf r})ds=0$, which
yields the following equation,

\begin{equation}
\frac{d}{ds}\left(n({\bf r})\frac{d{\bf r}}{ds}\right)=\nabla n({\bf
r}). \label{eqn:track}
\end{equation}

Generally, an optical ray is refracted in a nonhomogeneous medium,
but electrons do not. This motivate us to assume an energetic gluon
jet is refracted while a quark jet always goes straightly in the
non-homogeneous medium. On a microscopic level, lattice data
suggests~\cite{liaojf06} that there could be massive bound states in
strongly-couple QGP around the critical temperature.  These bound
states could serve as colored scattering centers for gluons, and
lead to refraction.


For a photon with energy $E$, the refractive index $n(E)$ is related
to the density of the scattering centers in medium N, $
n(E)=1+\frac{2\pi N}{E}F(E)$, where $F(E)$ is the forward scattering
amplitude~\cite{dremin}. Similarly, we assume the refractive index
for a gluon jet is related to the initial gluon density of the
medium,

\begin{eqnarray}
n(E, \tau, {\bf b}, {\bf r})=1+\kappa (E)\rho_g(\tau,{\bf b},{\bf
r}),\label{eqn:gluon-refrac}
\end{eqnarray}
where the $\kappa(E)$ factor should be related to the dispersive
power of a gluon jet with a given energy, similar to ${2\pi}F(E)/E$
in Ref.\cite{dremin}. Here for a simple calculation, we choose a
fixed value of $\kappa(E)$ by fitting data.

\begin{figure}[tb]
\includegraphics[width=0.9\linewidth]{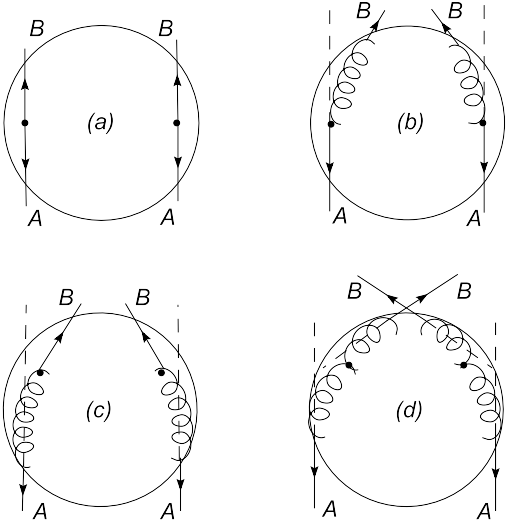}
\caption{\label{fig:diha-tangen} The trajectories of four types of
jet pairs for quark-quark (a), quark-gluon (b), gluon-quark (c) and
gluon-gluon (d), respectively, back-to-back created in central
nucleus-nucleus collisions. A and B represent the direction of
trigger and away-side, respectively.}
\end{figure}

Neglecting transverse expansion, the gluon density distribution in a
1D expanding medium in $A+A$ collisions at impact-parameter ${\bf
b}$ is assumed to be proportional to the number of binary
collisions,

 \begin{eqnarray}
 \label{eqn:glu_dens}
  \rho_g(\tau,{\bf b},{\bf r})&=&\frac{\tau_0\rho_0}{\tau}
  \frac{8\pi^2 R_A^4}{9A^2}\, t_A(|{\bf r}+{\bf n}\tau|)
 \nonumber\\
        &&\times~ t_A (|{\bf b}-({\bf r}+{\bf n}\tau)|)\, .
 \end{eqnarray}

A quark jet produced at ${\bf r}$ always hold its initial direction
${\bf n}$ while a gluon jet produced at ${\bf r}$ is deflected from
its initial direction in the medium before fragmenting into hadrons
outside the overlap region. $t_A(r)=\frac{3A}{2\pi
R^2}\sqrt{1-r^2/R^2}$ is the nuclear thickness function in a
hard-sphere geometry model. We use a set of parameters same as those
in Ref. \cite{zoww07}, $\mu_0=1.5$ GeV, $\epsilon_0\lambda_0=0.5$
GeV and $\tau_0=0.2$ fm/$c$ in Eq.
(\ref{eqn:enloss})(\ref{eqn:loss})(\ref{eqn:glu_dens}). The total
energy loss of a quark or a gluon jet is all given by Eq.
(\ref{eqn:enloss})(\ref{eqn:loss})(\ref{eqn:glu_dens}), except that
the gluon jet follows a curved path given by Eq.
(\ref{eqn:track})(\ref{eqn:gluon-refrac}). In the most central
$Au+Au$ collision, the refractive index can be simply written as
$n(r)=1+2\kappa(1-r^2/R^2)$. When we choose $\kappa =0.25$ and
energy loss parameter $\epsilon_0=1.7$ GeV/fm, numerical results for
single and dihadron spectra~\cite{zw08} can fit data well.

\begin{figure}[tb]
\includegraphics[width=0.9\linewidth]{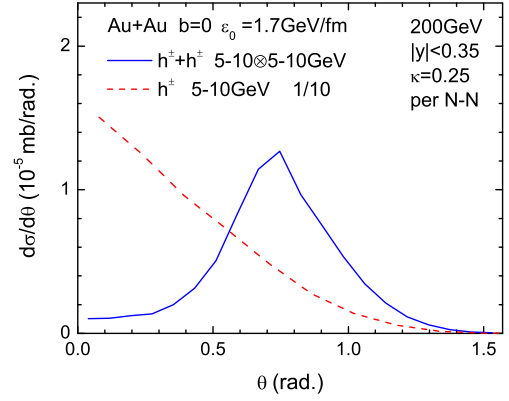}
\caption{\label{fig:singl-di-dangle} (color online). The single
hadron (dashing curve) and dihadron (solid curve) cross sections
only contributed respectively from single gluon jets and $qg/gq$ jet
pairs as a function of $\theta$ of gluon jets deflected from its
initial direction when the jets arrive at the system surface.}
\end{figure}

\begin{figure}[tb]
\includegraphics[width=0.9\linewidth]{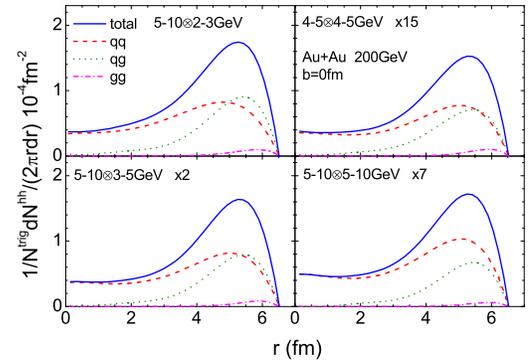}
\caption{\label{fig:dndr} (color online). The dihadron yield per
unit area as a function of initial spatial place $r$ of surviving
and fragmented jet pairs in the transverse plane in most central
$Au+Au$ collisions.}
\end{figure}

Figure~(\ref{fig:diha-tangen}) shows the trajectories for the four
possible hard-scattered jet pairs, i.e. quark-quark (a), quark-gluon
(b), gluon-quark (c) and gluon-gluon (d). The refraction results in
a deflection of the gluon, thus the quark-gluon and gluon-gluon jet
pairs are no longer back-to-back. Consequently, this a acoplanarity
propagates to the away-side hadron pairs after fragmentation. In
addition, the gluon deflects towards higher density region, leading
to a longer path and more energy loss for a gluon jet, so its
contribution to observed hadron pairs is reduced.
Figure~(\ref{fig:singl-di-dangle}) shows the distribution of the
deflection angle $\theta$, for single hadron from gluons and hadron
pairs from qg/gq di-jets. The single hadron distribution peaks at
$\theta=0$, but has a long tail, this is consistent with the
surface emission picture, where the observed jets traverse short
path in the medium, thus have small deflection angles. However, the
dihadron from quark-gluon dijet peaks around $\theta_c=0.72$,
implying that gluon jet selected by the hadron pair has to go
through on average a finite medium, thus suffers a substantial
deflection.

Figure~(\ref{fig:dndr}) shows the distribution of the origin the
jet pairs contributing to the observed dihadrons as a function of
the initial spatial place $r$ in the transverse plane for b=0fm.
Since the energy loss of a gluon jet is about 2 times of that of a
quark jet, gluon-gluon contributions are  greatly suppressed and
can be ignored in the most central $Au+Au$ collisions. The
quark-gluon contributions come dominatingly from the outer layer of
medium. The dihadron yields in
Figure~(\ref{fig:singl-di-dangle})(\ref{fig:dndr}) have been
averaged over the azimuthal angle of hadron pairs. Because a gluon
jet has a biggest deflected angle if created with a moving
direction perpendicular to the gradient of the index of refraction,
the descriptions for $q-g$ contributions in
Figure~(\ref{fig:singl-di-dangle})(\ref{fig:dndr}) are consistent
to the tangential surface emission bias.

\begin{figure}[tb]
\includegraphics[width=1.0\linewidth]{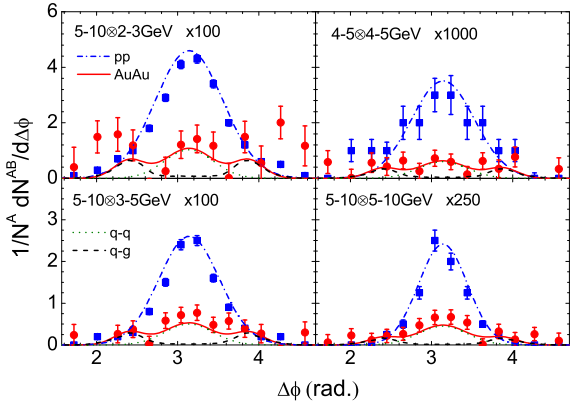}
\caption{\label{fig:phi-diha} (color online). Per-trigger yield
versus $\Delta\phi$ for various trigger and partner $p_{T}$
($p_T^{A}\otimes p_T^{B}$), arranged by increasing pair momentum
(sum of $p_T^{A}$ and $p_{T}^{B}$), in $p+p$ and most central
$Au+Au$ collisions. The Data are from Ref. \cite{phenix-jjia07}.}
\end{figure}

The NLO framework of Ref. \cite{zw08} is used to calculate the
away-side dihadron multiplicity separately for $q-q$, $q-g$ and
$g-g$ process. However, the azimuthal distribution of dihadron in a
NLO pQCD is divergent at $\Delta\phi=\pi$ since there are large
logarithmic high order corrections (Sudakov logs) which need be
resummed \cite{owens}. Such divergence is generally treated by
smearing $\Delta\phi$ to match the p+p shape~\cite{xnw04} but keep
the integral over $\Delta\phi$ fixed. In this paper, we follow the
same approach for the case of $p+p$ collisions, and for quark-quark
contributions in $Au+Au$ collisions by simply smearing around
$\Delta\phi=\pi$. Because of gluon refraction in $Au+Au$ collisions,
the smearing technique isn't needed for quark-gluon and gluon-gluon
contributions, as shown in Figure (\ref{fig:singl-di-dangle}) where
the Gaussian curve is directly calculated out without any smearing.

The final per-trigger yield $\Delta\phi$ distribution are shown in
Figure (\ref{fig:phi-diha}). The calculation is compared with
PHENIX data in several trigger ``A'' and partner ``B'' $p_{T}$
($p_T^{A}\otimes p_T^{B}$) bins. Our calculation clearly shows one
center peak around $\Delta\phi=\pi$ and two side peaks around
$\Delta\phi=\pi\pm0.72$ at the away-side. The center peak (``head''
region) is domianted by the quark-quark jet pairs, while the two
side peaks (``shoulder'' region) are dominated by quark-gluon and
gluon-gluon dijets.
However for the $5-10 \otimes 5-10$ GeV/$c$ bin, the side peak
amplitude decreases relative to the center peak. This is because
gluon jet contributions to hadrons decrease with increasing hadron
$p_T$, thus the dihadron yield is dominated by the quark-quark jet
pairs. This effect can also be seen in Figure (\ref{fig:dndr}). Our
side-peak location of 0.72 is still smaller than the experimental
value of 1.1, suggesting there might be additional deflection
mechanism.


We can also quantify the relative importance of the side peaks and
center peak using the $R_{HS}$ introduced by
PHENIX~\cite{phenix-jjia07}. In our case, it simply reflects the
relative contribution from quark-gluon process and quark-quark
process (neglecting contributions from gluon-gluon), i.e,

\begin{eqnarray}
R_{HS}\approx\frac{\int d\Delta\phi(dN^{AB}/d\Delta\phi)^{qq}} {\int
d\Delta\phi(dN^{AB}/d\Delta\phi)^{qg+gq}}=R_{qq/qg} \label{eqn:R_qqoverother}
\end{eqnarray}

\begin{figure}[tb]
\includegraphics[width=1.0\linewidth]{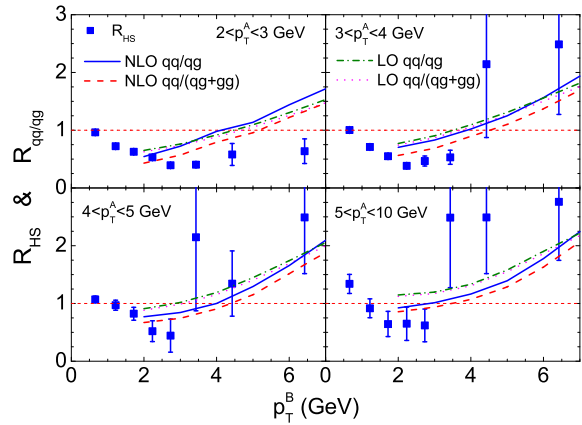}
\caption{\label{fig:R_HS} (color online). Comparisons of
$R_{qq/(qg+gg)}$ and $R_{HS}$ as functions of the transverse
momentum of the associated hadron with increasing trigger bins in
$0-20\%$ $Au+Au$ collisions at 200 GeV. The Data are from Ref.
\cite{phenix-jjia07}.}
\end{figure}

$R_{qq/qg}$ is insensitive to $\kappa(E)$ and $n(E, \tau, {\bf b},
{\bf r})$ according to  Eq. (\ref{eqn:gluon-refrac}). Figure
(\ref{fig:R_HS}) shows the comparison of the  $R_{qq/qg}$ with the
experimental data. The excellent agreement suggests that gluon
deflection is reasonable. And the $R_{HS}$ largely reflects the
ratio of the surviving quark-gluon to quark-quark in central
collisions.

We have repeated similar calculations for LHC energy. Since both
gluon-gluon and quark-gluon dominates over quark-quark up to
$p_T\sim100$ GeV/$c$, we expect $R_{HS}<1$. Since gluon-gluon has
stronger deflection than quark-gluon, there might even exist
multiple side peaks~\cite{zw08}.

In summary, we argue that the refraction for gluon jet might be a
possible mechanism for the rich and $p_T$ dependent away-side shape
in central Au+Au collisions observed at RHIC. The gluon refraction
is implemented in the NLO pQCD energy loss framework to obtain the
away-side $\Delta\phi$ distribution. The away-side shows a center
peak in the head region and two side-peaks in the shoulder region.
The former is dominated by the quark-quark dijets while the latter
is dominated by the $qg/gq$ dijets. The different proportions of
contributions from $qq$, $qg/gq$ and $gg$ pairs determine whether
there are one, or three peaks in dihadron $\Delta\phi$ distribution.
Our calculation qualitatively describes the experimental findings,
however the refraction angle under-predicts the split angle seen in
the data. This might be due to simplification in our assumption or
additional mechanisms need to be included. If gluon jets are indeed
refracted, one might expected different shape at LHC at the same
$p_T$. Also since the high $p_T$ (anti-)proton production is
dominated by hard gluon fragmentation, the azimuthal angle
distribution of the high $p_T$ (anti-)proton-(anti-)proton pairs
might have different shape from that of the charged hadron pairs
measured at RHIC. These studies will be done in our future works.


We thank Jiarong Li, Xin-Nian Wang and Jiangyong Jia for helpful
discussions, and especially thank J. F. Owens for discussing and
providing his NLO codes. This work was supported by MOE of China
under Projects No. IRT0624, No. NCET-04-0744 and No.
SRFDP-20040511005, and by NSFC of China under Projects No.
10440420018, No. 10475031 and No. 10635020.


\end{document}